# Characterisation of CMOS Image Sensor Performance in Low Light Automotive Applications

Shane P. Gilroy[1,2], John O'Dwyer[2] and Lucas C. Bortoleto[2]

[1]*Department of Mechanical and Electronic Engineering, Institute of Technology Sligo, Co. Sligo, Ireland*
[2]*Department of Engineering Technology, Waterford Institute of Technology, Co. Waterford, Ireland*

**Abstract**

The applications of automotive cameras in Advanced Driver-Assistance Systems (ADAS) are growing rapidly as automotive manufacturers strive to provide 360˚ protection for their customers. Vision systems must capture high quality images in both daytime and night-time scenarios in order to produce the large informational content required for software analysis in applications such as lane departure, pedestrian detection and collision detection. The challenge in producing high quality images in low light scenarios is that the signal to noise ratio is greatly reduced. This can result in noise becoming the dominant factor in a captured image thereby making these safety systems less effective at night. This paper outlines a systematic method for characterisation of state of the art image sensor performance in response to noise, so as to improve the design and performance of automotive cameras in low light scenarios. The experiment outlined in this paper demonstrates how this method can be used to characterise the performance of CMOS image sensors in response to electrical noise on the power lines**.**

## 1  Introduction

There were 472 road deaths in Ireland in 1997 according to the Road Safety Authority [RSA, 2015]. There were 166 road deaths in 2015 and this number of fatalities has decreased in an almost linear fashion in the intervening years as shown in Figure 1. This reduction is due in part to the advancement and standardisation of automotive safety systems. Automotive safety systems fall into two main categories, passive safety systems which protect the driver and passengers in the event of a collision such as seat belts, airbags etc. and active safety systems which prevent the occurrence of a collision such as traction control, ABS Brakes and blind spot monitoring. Intelligent vision systems are quickly becoming a larger component of active automotive safety systems for a wide range of applications as manufacturers strive to provide 360˚ protection for their customers. One such example of a vision based safety system is lane departure prevention. Failure to stay in the correct lane was the largest single factor in fatal collisions in the US in 2002 [NHTSA, 2003], playing a role in 32.8% of all cases. Vision Systems can be used to detect lane departure before a collision occurs and mitigate the risk by alerting the diver, adjusting the steering angle or applying the brakes on the opposite side of the vehicle in order to prevent lane departure.

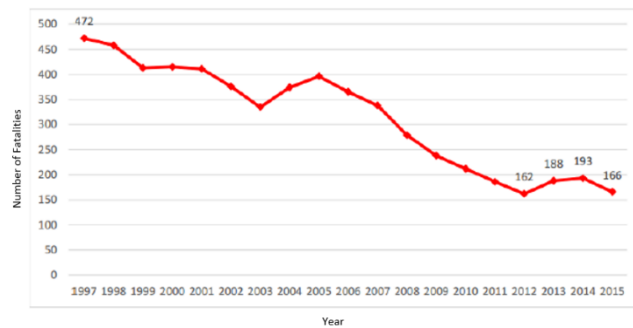

Figure 1: Road Deaths Ireland 1997-2015. Almost linear decline in road deaths in modern times as automotive safety systems improve.

Another example of a vision based safety system is backover protection. There is an average of 232 fatalities and over 13,000 injuries each year in the US as a result of backovers [Naylor, 2014, Singh S, 2014]. The victims are primarily children under the age of 5 years old and the elderly. As a result, new US legislation will be enforced from May 2018 requiring the mandatory installation of rear view cameras on all new vehicles weighing under 4,500kg. This law will apply to all cars, SUVs, buses and light trucks. A minimum field of view of a 3m by 6m zone directly behind the vehicle and minimum image size is specified. These cameras will not only reduce the risk of backovers by allowing the user to see behind the vehicle, but also through the use of advanced software features such as Object Detection which can allow the vehicle to mitigate the risk of backovers by autonomously alerting the user or by applying emergency braking etc.

In order to be effective in safety applications, these systems must perform without error in both daytime and night time scenarios. In low light situations, the Signal to Noise Ratio (SNR) of a captured image can be greatly reduced. This can lead to noise becoming the dominant signal, as in Figure 2, reducing the amount of useful information available for software analysis thereby impacting the performance of the safety system or in extreme cases making the safety system defunct. In addition to this, technological trends have shown a demand for increased spatial resolution in image sensing applications. In order to obtain increased resolution without increasing the physical size of the sensor, pixel size must be reduced [Gamal, 2009, Gao, Yao et al., 2013]. This further reduces the amount of light captured by each individual pixel and contributes to the reduction of Signal to Noise Ratio in low light applications.

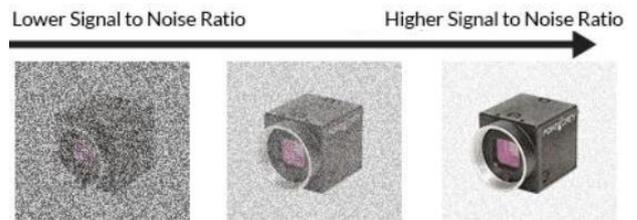

Figure 2: Effect of Signal to Noise Ratio (SNR) on Image Quality. A reduced SNR can result in less analysable content making vision based safety system defunct in low light applications.

## 2     Current State of the Art and Knowledge Gap

Current research such as Seo et al 2013 [Seo, Sawamoto et al., 2013], Chen et al 2012 [Chen, Xu et al., 2012] and Feruglio et al 2006 [Feruglio, Pinna et al., 2006] all attempt to address this issue at image sensor chip level with a view to reducing internal noise such as temporal, fixed pattern, shot and read noise. These chip level improvements can be very difficult to observe in practical automotive applications as the dominant sources of noise are often created external to the image sensor from the surrounding circuitry and the automotive environment. This can result in camera designers not achieving the full capability of the image sensors low light performance. A knowledge gap exists for a systematic method to analyse how individual state of the art image sensors perform in response to the application specific sources of noise that may occur within the camera unit and particularly within the automotive vehicle.

## 3     Hypothesis

The method outlined in this paper can be used to systematically characterise modern image sensor performance in response to injected noise as desired by the designer. This allows engineers to tailor schematic and PCB design to filter the precise critical ranges unique to each new model of image sensor in order to maximize the low light performance of automotive cameras.

### 3.1 Methodology

A method has been developed to characterise CMOS image sensor performance in response to electrical noise on the power supply lines, Figure 3. Image sensor characterisation is carried out by setting the image sensor to stream video data and covering the lens with a non-transparent material in order to ensure that no light strikes any part of the image sensor throughout the characterisation process. Electrical noise of a specific frequency is then coupled on to the image sensor power supply lines using a noise generator. RAW images are captured from the image sensor and converted into RGB form for analysis using a custom row noise algorithm developed in Matlab in order to quantify image noise. The noise frequency step can be increased and the process repeated for the desired frequency range allowing systematic characterisation of the image sensor performance in response to power supply noise.

The proposed characterisation method is automated with the use of LabVIEW and National Instruments TestStand in order to manage the operation of this analysis for the thousands of frequency steps required when conducting characterisation over a wide frequency spectrum [Gilroy, 2016].

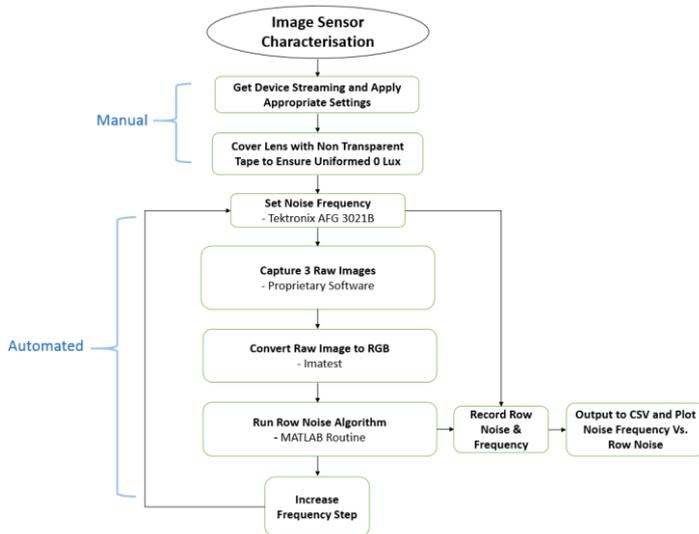

Figure 3: Image Sensor Characterisation Flowchart

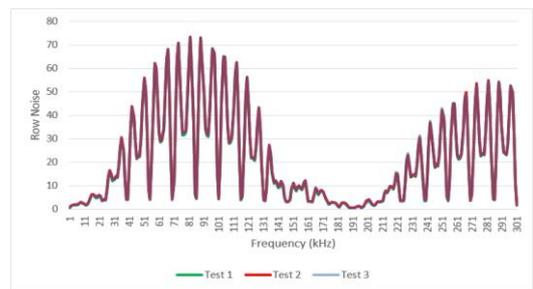

Figure 4: Characterisation Repeatability. Ominvision OV7955 image sensor performance characterisation in reponse to electrical noise carried out 3 times and overlaid to demonstrate repeatability.

Repeatability of the characterisation method has been confirmed using an Omnivision OV7955 image sensor. Image sensor characterisation was repeated three times in succession for a frequency input range of 50Hz to 300 kHz, Figure 4. The test setup has been dismantled and reinstalled between the second and the third test run to rule out any dependency on setup. The results indicate that the characterisation method is highly repeatable.

## 4   Experimental Results

Performance characterisation in response to electrical noise on the power supply lines has been carried out on two state-of-the-art image sensor models in order to display the use of the methodology proposed in this article. The results of this characterisation can be seen in Figure 5.

Detailed information can be derived from the characterisation process to identify critical ranges of power supply noise that an individual image sensor model is particularly susceptible to. This information can be used in the design of tailored hardware and software filters to reduce the impact of noise on a specific CMOS image sensor and maximise the performance of vision systems for safety critical applications at low light [Gilroy, 2016].

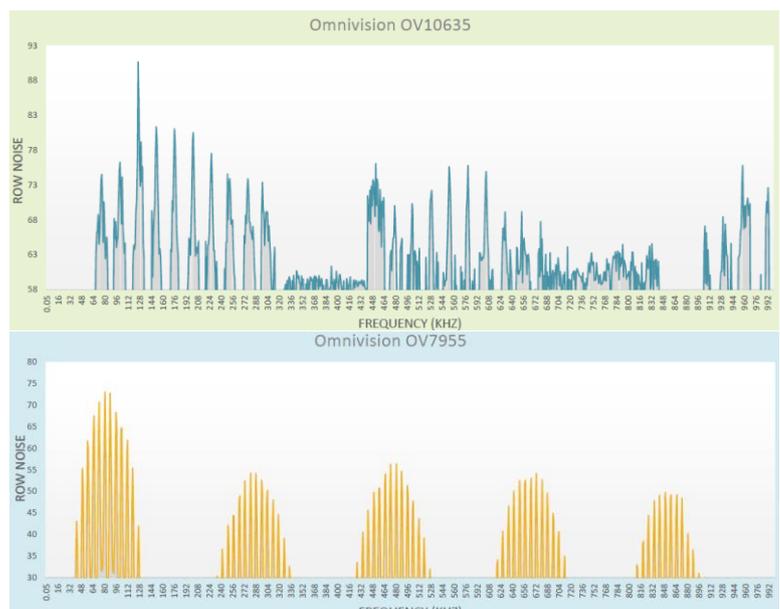

Figure 5: Image sensor performance characterisation in response to electrical noise conducted on to the power supply lines 0Hz-100kHz

# 5  Conclusions

A systematic, repeatable characterisation method of image sensor performance in response to power supply noise has been proposed which can be used to identify the specific noise frequency ranges that each individual model of image sensor is immune or susceptible to across a wide frequency spectrum. Characterisation of two state-of-the-art image sensor models has been carried out in order to demonstrate the use and effectiveness of the proposed method. The results provided by the characterisation method can be used to identify peak impact and critical ranges of application specific noise that can be used as a design input for the component selection, schematic, PCB and software design of vision systems to improve low light performance in critical safety applications. The proposed characterisation method has focused on image sensor performance in response to electrical noise on the power supply lines only, however the method can be adapted in future to conduct image sensor characterisation in response to any noise source or stimulus as required by the vision system application. The structured, systematic nature of the proposed characterisation method allows the system to be updated by switching the custom row noise algorithm executable file with one defined by the new noise input allowing characterisation to be implemented by following the same steps. The method outlined could also be modified in future to allow the characterisation of complete camera modules for the purposes of debug or validation of vision systems in response to electrical noise.

## Acknowledgements

This research was carried out in association with Valeo Vision Systems, Tuam, Co. Galway, Ireland.